\documentclass[english,aps,prb,showpacs,twocolumn]{revtex4}
\usepackage{graphicx}
\usepackage{babel}

\begin{document}

\title{Bound-to-bound and bound-to-continuum optical transitions in combined quantum dot - superlattice systems}

\author{F.F.~Schrey}

\author{L.~Rebohle}

\author{T.~M\"uller}

\author{G.~Strasser}

\author{K.~Unterrainer}

\affiliation{Institut f\"ur Photonik und Festk\"orperelektronik, Technische Universit\"at Wien,
Floragasse 7, A-1040 Wien, Austria}

\author{D.P.~Nguyen}

\author{N.~Regnault}

\author{R.~Ferreira}

\author{G.~Bastard}

\affiliation{Laboratoire Pierre Agrain - Ecole Normale Sup\'erieure,
24 rue Lhomond, F-75005 Paris, France}

\begin{abstract}

By combining band gap engineering with the self-organized growth of quantum dots, we present a scheme of adjusting the mid-infrared absorption properties to desired energy transitions in quantum dot based photodetectors. Embedding the self organized InAs quantum dots into an AlAs/GaAs superlattice enables us to tune the optical transition energy by changing the superlattice period as well as by changing the growth conditions of the dots. Using a one band envelope function framework we are able, in a fully three dimensional calculation, to predict the photocurrent spectra of these devices as well as their polarization properties. The calculations further predict a strong impact of the dots on the superlattices minibands. The impact of vertical dot alignment or misalignment on the absorption properties of this dot/superlattice structure is investigated. The observed photocurrent spectra of vertically coupled quantum dot stacks show very good agreement with the calculations.In these experiments, vertically coupled quantum dot stacks show the best performance in the desired photodetector application.
\end{abstract}


\maketitle

\section{Introduction}

InAs quantum dots (QDs) embedded in GaAs quantum well structures have initiated considerable research activities in the recent years \cite{Chu991, Horiguchi99, Liu01, Tang01, Hofer01, Rebohle02, Maimon98, Sauvage01, Ye02, Chu992, Kim02, Berryman97, Chen02, Maximov99}. The interest in QDs is mainly based on the favorable energy spacing of their bound electronic states and the great potential to adjust these properties. For the use in infrared (IR) devices, the electron level structure covers the important spectral region between some 40~meV up to 400~meV. Although quantum well structures were already successfully used for infrared (IR) light detection \cite{Levine93}, these structures suffer from their insensitivity to normal light incidence. In contrast, QDs are able to absorb efficiently IR light under normal incidence \cite{Finkman01, Kim04, Hirakawa99}, which considerably simplifies the layout of potential photodetectors and sensor applications. Furthermore, QD structures are expected to provide higher photocurrents and lower dark currents than quantum well structures due to the longer lifetime of the excited states \cite{Bockelmann90, Urayama01}. 

The IR photoresponse of InAs dots embedded in GaAs was investigated very recently \cite{Chu991, Horiguchi99, Liu01, Tang01, Hofer01, Rebohle02, Maimon98, Sauvage01, Ye02} and transition energies between the dot ground state and the GaAs conduction band were found in the 100-400~meV range. Depending on the dot potential shape and the doping level, photodetection can even be extended into the region below 100~meV \cite{Phillips98, Aslan03}. Furthermore the spectral response of such devices can be controlled through the applied bias allowing multiwavelength detection \cite{Zhengmao02}. It was also shown, that the electronic states within the QD can be tuned by changing the InAs dot size \cite{Chu992} by incorporating an AlAs layer close to the QDs \cite{Kim02}, or by confining the QDs in an ${\rm Al}_{\rm x}{\rm Ga}_{1-{\rm x}}{\rm As}$ Matrix \cite{Berryman97, Vasanelli01}, which causes a higher conduction band offset to InAs.

In this work we design the energy level scheme of QD structures by combining band gap engineering with the self-organized growth of QDs in a superlattice structure (SL). A simple approach to the design of the device by assuming that the SL can be treated independently from the dots leads to reasonable predictions of the transition energies between the InAs QD ground state and the continuum states of the surrounding GaAs/AlAs system. Nevertheless, this approach by solving one-dimensional Schr\"odinger equations fails completely for predictions about bound-to-bound transitions and possible polarization dependencies of the absorption spectra. Compared to regular dots, the dots inserted inside the AlAs barrier should exhibit  pronounced changes for the delocalized states, while the eigenstates, which are well localized in the dots should be less or much less affected by the AlAs barriers. In fact, since the InAs dot represents a deep potential barrier which at the same time extends over a non negligible fraction of the SL unit cell, one cannot expect to describe accurately the continuum states of the dot by means of a simple 3D GaAs continuum. The latter would be size-quantized into 2D subbands by the introduction of the AlAs barrier. Hence, it becomes mandatory to make a three dimensional computation of the continuum states of the dot, if one wants to understand their photoconductive properties. We will show that the optical transitions within these detectors can be tuned by changing either the dot growth conditions or the SL parameters. In addition, we will compare vertically aligned QD stacks with those of isolated single or double QDs in these SL structures, in order to better understand the influence of the dots on the SL. Furthermore, we will demonstrate that the AlAs barriers of the SL reduce the dark current considerably, which is of great importance regarding the application as photodetector.

\section{Sample Preparation and Setup}

\begin{figure}[!htbp]
\begin{center}\includegraphics[width=8.5cm,
  keepaspectratio]{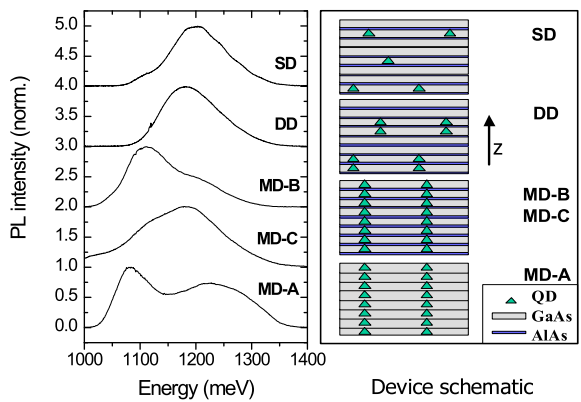}
\end{center}
\caption{Interband photoluminescence spectra of the investigated devices. On the right: Sketches of the
different device layouts. The arrow denotes z, the growth direction.} \label{fig:device}
\end{figure}

All samples in the following experiments were grown by molecular beam epitaxy on smooth semi-insulating GaAs (001) substrates. After depositing a 650~nm thick Si-doped GaAs layer the SL structure is formed, whereby the InAs dots were grown by the Stranski-Krastanov mechanism by depositing 2 monolayers of InAs. A Si-doped GaAs top layer covers the devices. Whereas the bottom contact layer was grown at $600^\circ{\rm C}$, all subsequent layers were deposited at $485^\circ{\rm C}$ in order to avoid Al, Ga and In intermixing from the AlAs barriers, GaAs wells, and InAs dots. We obtain a lateral QD density of approximately $5\times 10^{10} {\rm dots}/{\rm cm}^2$ per layer. The samples differ in the number of QD-layers, the vertical layer spacing or the AlAs/GaAs SL period respectively. An overview over the sample parameters is given in table \ref{tab:device}, while the basic device design is sketched in fig. \ref{fig:device}.

\begin{table}
\caption{\label{tab:device} devices}
\begin{ruledtabular}
\begin{tabular}{lccccc}
Device & MD-A & MD-B & MD-C & DD & SD\\
SL period & 10~nm & 11~nm & 14~nm &  11~nm & 11~nm\\
Number of QD layers &  30 & 20 & 20 & 20 & 20 \\
Number of AlAs barriers & - & 20 & 20 & 40 & 60
\end{tabular}
\end{ruledtabular}
\end{table}

The first set of samples, three multiple dot (MD) device structures, in the following denoted as MD-A, MD-B and MD-C, were designed, grown and characterized. Device MD-A consists of periodically arranged InAs QD layers, which are spaced by a 10~nm thick GaAs matrix. As shown in a previous work \cite{Hofer01}, the small separation causes a vertical alignment of the QDs due to their strain distribution. Device MD-B is additionally provided with 1~nm thick AlAs barriers at a distance of 1~nm from the QD layer resulting in a SL period of 11~nm. We expect that due to the periodic AlAs barriers the intersubband absorption will be blueshifted compared to MD-A. Since the QD growth conditions between sample MD-A and MD-B were changed to obtain a higher transition energy of the ground state emission, we decided to design a third device MD-C, in which the main absorption peak is tuned to an energy comparable to the transition of sample MD-A. Thus the SL period was increased to 14~nm. 

The second set of samples, a single dot (SD) and a vertical double dot (DD) device are fabricated by growing one (SD) or two (DD) SL periods under the same conditions as device MD-B. These layers are followed by 2 SL periods without QD layers preventing the vertical alignment of the QDs. This sequence is repeated several times in such a way that the total number of QD layers in all devices is 20 or 30 (only MD-A). 

To study the energy level scheme and the dark current behavior of the nanostructures photoluminescence (PL), photocurrent (PC) and current-voltage (IV) measurements were performed by mounting the devices onto the cold finger of a He flow cryostat system. The PL investigations were carried out by HeNe laser excitation at  632.8~nm, a lock-in amplifier and an InGaAs photodiode. For the PC measurements the devices were processed by photolithography and wet chemical etching into mesas. Furthermore, the cleaved edges of the sample were polished in order to obtain a waveguide structure below the mesa. Finally, the devices were provided with back and front contacts made of a thermally alloyed Ni/Ge/Au layer. The spectral dependence of the photoresponse of the devices is measured in TE as well as TM geometry using a standard FTIR spectrometer with a glow-bar infrared source and a mid-infrared polarizer. The I-V characteristics of the mesa structures were recorded with a HP 4155A semiconductor parameter analyzer.

\section{Model}

\begin{figure}[!htbp]
\begin{center}\includegraphics[width=8cm,
  keepaspectratio]{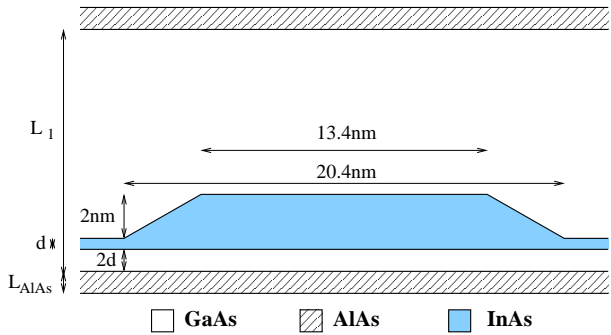}
\end{center}
\caption{schematic representation of the supercell including the dot and its WL. $L_1=10$~nm (MD-A, MD-B) or 13~nm (MD-C) and $L_{AlAs}=0$ (MD-A) or 1~nm (MD-B, MD-C).}
\label{dotmodel}
\end{figure}

We used a one band envelope function framework where the conduction band effective mass is $0.07 m_0$.  This scheme has proven to be successful in interpreting the intricated magneto-optical spectra associated with the bound-to-bound transitions in the "free" InAs dots \cite{Isaia02} including the magneto-polaron effect.  The conduction band discontinuity between AlAs and GaAs was taken equal 1.08~eV \cite{Ferreira91} and that between GaAs and InAs to 0.4~eV.  The dot shape was approximated by a truncated cone making an angle of $30^\circ$ with the dot basis (see fig. \ref{dotmodel}).  The lower (upper) radius of the cone is 10.2~nm (6.7~nm).  The dot stays on a $d=0.565$~nm thick wetting layer (WL).  The dot height is taken equal to 2~nm.  With such values the energy splitting between the two lowest lying states ($S$ and $P^\pm$) is about 50~meV and the energy distance from the ground bound state and the edge of the continuum is about 160~meV.  Note that the latter is mainly determined by the vertical dimension of the dot while the $S-P$ splitting is mostly governed by the lateral dimensions of the dot.

The numerical method consists of a block diagonalization of the dot Hamiltonian on a large basis of 130.000 functions \footnote{source code is available at http://www.phys.ens.fr/$\sim$regnault/diagham}. The basis consists of plane waves, which are made periodic with the periodicity of a supercell\cite{Lelong96}.  The dimensions of the supercell are 91~nm along the $x$, $y$ (in-plane) directions and $L=L_1+L_{AlAs}$ along the growth ($z$) directions (see figure \ref{dotmodel}).  The supercell is chunked into elementary cubic cells (cube side: $d=0.565$~nm).  Within each such cell the potential is constant.  The first two hundreds eigenstates are evaluated by means of an exact diagonalization which uses the L\'anczos algorithm \cite{Nguyen04}.  Since translation symmetry in the growth direction exists in this model, the wave vector $k_{z}$ is a good quantum number.

\begin{figure}[!htbp]
\begin{center}\includegraphics[width=8.5cm,
  keepaspectratio]{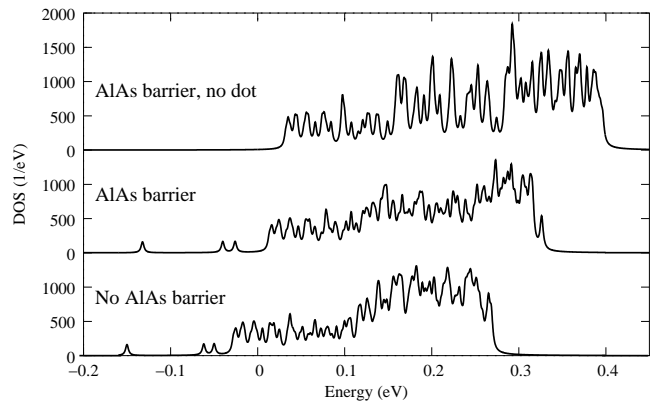}
\end{center}
\caption{Density of states of even-even symmetry (even wave function in $x$ and $y$) for electron with $L=12$~nm (for $k_z=0$)}
\label{DOS_S}
\end{figure}

A comparison between the calculated density of states whose wave functions are even in $x$ and $y$ directions for a single quantum dot with or without AlAs barrier is shown in fig. \ref{DOS_S} for a dot plane located at $\simeq$2d from the AlAs barrier.  In the calculation for the dot sample without AlAs barriers the latter was replaced by a GaAs layer with the same thickness. In practice, we broadened the delta function by replacing it by a Lorentzian of 8~meV full width at half maximum.  The high energy cutoffs in fig. \ref{DOS_S} are unphysical and arise from our restriction to the lowest two hundreds eigenvalues of the conduction states.  We see in fig. \ref{DOS_S} that the AlAs size quantization affects the eigenstates quite differently depending on their nature: the states which are bound to the dots are much less affected than the WL states since the $z$ extension of the latter states is larger than those of the formers.  This implies that the absorption edge for bound-to-continuum transitions is blue shifted by the extra confinement brought by the AlAs layers. The absence of QDs in the SL shifts the onset of the first miniband to higher energies above the GaAs conduction band. This underlines as well the strong impact of the dots on the energetic structure of the SL.

To take advantage of the rotation symmetry in the plane existing in our dots, we then performed calculations on a basis of Fourier-Bessel functions:
\begin{equation}
\psi_{lnq}(\overrightarrow{r}) = C_{lnq} \cdot \exp (i l\theta) \cdot J_{l} \big (\lambda_{n} ^{l} \frac{r}{R}\big ) \cdot \exp \left(i \frac{2 \pi q z}{L}\right) 
\end{equation}

where $C_{lnq}$ is a normalization constant, $l$ is the angular quantum number, $\lambda_{n} ^{l}$ are zeros of corresponding Bessel functions of the first kind $J_{l}$ and $q$ an integer. The height and radius of super-cylinder are $L=L_1+L_{AlAs}$ and $R=200$~nm respectively. In this basis, the eigenstates are classified according to the $z$ projection of the electron angular momentum: S ($l = 0$), P ($l = 1$), D ($l = 2$), ... The optical selection rules \cite{Bastard96} state that the transitions from the ground state, which is of symmetry S, to S symmetry states contribute to absorption with linear polarization in $z$ direction while only the transitions to P states give rise to absorption of in-plane polarizations. In the calculation of the absorption spectrum, one must add the contribution of the first Brillouin zone $k_z\in\left]-\frac{\pi}{L}, \frac{\pi}{L}\right]$.

\begin{figure}[!htbp]
\begin{center}\includegraphics[width=8.5cm,
  keepaspectratio]{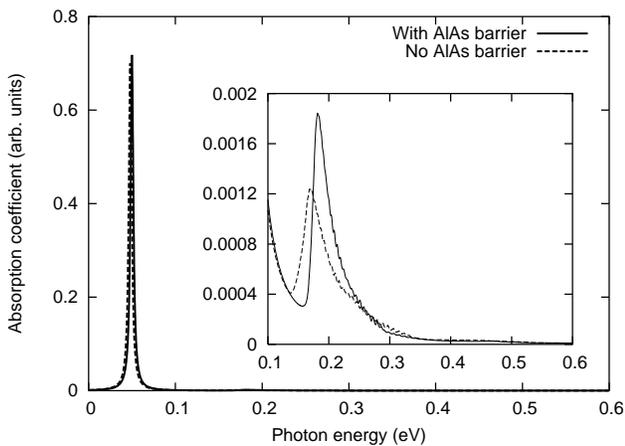}
\end{center}
\caption{Absorption coefficient for in-plane polarization with $L_1=10$~nm (MD-A and MD-B). The inset shows a zoom of the bound-to-continuum absorption.}
\label{polaxy}
\end{figure}

\begin{figure}[!htbp]
\begin{center}\includegraphics[width=8.5cm,
  keepaspectratio]{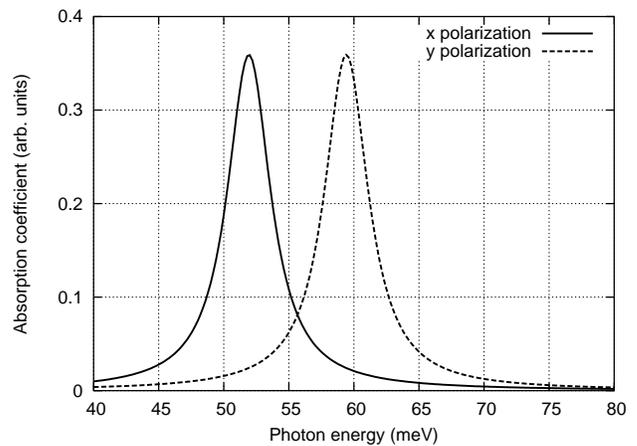}
\end{center}
\caption{Absorption coefficients for $x$ and $y$-oriented polarizations with $L_1=10$~nm (MD-B), AlAs barrier and a slightly elliptical dot $(R_y/R_x=0.9)$.}
\label{polaxyelliptical}
\end{figure}

\begin{figure}[!htbp]
\begin{center}\includegraphics[width=8.5cm,
  keepaspectratio]{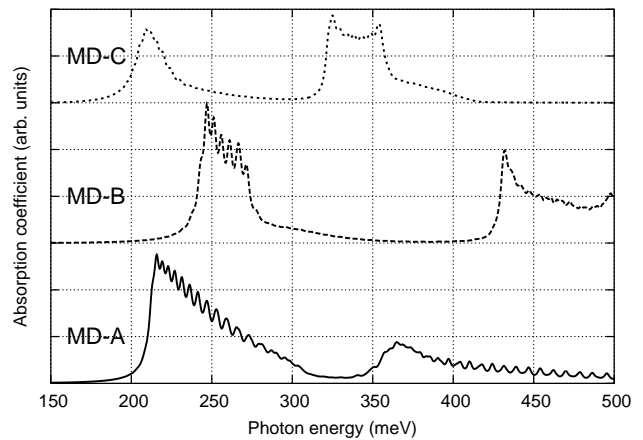}
\end{center}
\caption{Absorption coefficient for a $z$-oriented polarization. Full line: $L=10$~nm (MD-A). Dashed line: $L=11$~nm (MD-B). Dotted line: $L=14$~nm (MD-C). The same scale for the $y$-axis is used for the three samples.}
\label{polaz}
\end{figure}

Once the eigenvalues and eigenfunctions known, we compute the absorption coefficient $\alpha\left(\omega\right)$ as proportional to:
\begin{equation}
\alpha\left(\omega\right)\propto\sum_{n}\left|\langle s|\overrightarrow{\cal{E}}.\overrightarrow{r}| n\rangle\right|^2\delta\left(\epsilon_n-\epsilon_s-\omega\right)\hbar\omega\label{absorption} 
\end{equation}
where $n$ labels the discretized continuum of states, $\epsilon_n$ the corresponding eigenenergies and $\overrightarrow{\cal{E}}$ the light polarization.  There is a pronounced anisotropy of the quantum dot optical absorption (see fig. \ref{polaxy} and fig. \ref{polaz}). Small asymmetry of the dot produces a double peak structure for in-plane absorption coefficient (fig. \ref{polaxyelliptical}, see section IV for comparison with the experimental results). When the e-m wave is polarized along the $x$ (or $y$) direction, there exists a very strong absorption due to bound-to-bound transitions, the equivalent of the $S-P$ atomic absorption.  Consequently, the bound-to-continuum absorption is very weak. This is a result of the concentration of oscillator strength in the bound-to-bound transitions.  In the explored energy range it is essentially featureless: the bound-to-continuum absorption is vanishingly weak at the absorption onset as a result of the continuum states repulsion by the dot. This repulsion decreases smoothly with increasing kinetic energy since the dot parameters are such that there are no resonant states in the explored energy window with significant oscillator strength in this polarization.  The decreased repulsion leads to an absorption increase.  However, the absorption falls at high enough energy because of the decrease of the dipole matrix element, which is now associated with the destructive interference inside the dot of the initial and final states.  Very qualitatively speaking, taking the final states as plane waves, we see that the absorption will go to zero at large photon energy since the energy conservation forces the final state wavevector $\overrightarrow{k}_{fin}$ to be increasingly large while the dipole matrix element being the Fourier component of $\overrightarrow{r}\Psi_s\left(\overrightarrow{r}\right)$ at $\overrightarrow{k}_{fin}$ and therefore goes quickly to zero if $\overrightarrow{k}_{fin}l_{dot}\gg1$, where $l_{dot}$ is a typical length scale of $\overrightarrow{r}\Psi_s\left(\overrightarrow{r}\right)$, in practice the dot size \cite{Vasanelli01}. 
 
The optical absorption spectrum in the $z$ polarization for the three MD structures are shown in figure \ref{polaz}. The saw-tooth comes from the sum over a regular sample of the first Brillouin zone ($\simeq 50\;k_z$ values). This absorption is very different from the one of in-plane polarization and, in addition, shows a pronounced effect of the AlAs barriers. First, there is no bound-to-bound transition for the investigated dot parameters.  This is because the dots are flat objects, which support only bound states with no node along the growth direction, in order to minimize the kinetic energy.  This implies the existence of resonances in the continuum with nodes along $z$ inside the dot.  For the chosen dot parameters, the lower of these resonances falls in the investigated energy window. The $z$-dependence of its wavefunction resembles a sine function in the dot, granted that the bound state resembles a cosine function inside the dot.  Thus one may expect (and does calculate) a strong absorption for photon energy equal to the energy difference between this resonance and the ground state (provided that the resonance is not too blurred by its interaction with the continuum).  This is what happens both with and without AlAs barriers, as evidenced by the low energy peak in the A, B and C spectra. 

\begin{figure}[!htbp]
\begin{center}\includegraphics[width=8.5cm,
  keepaspectratio]{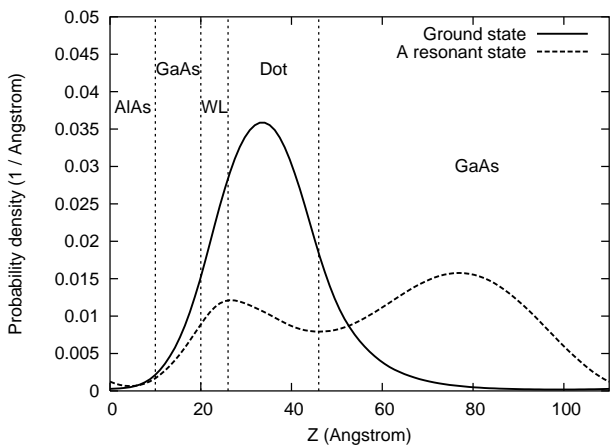}
\end{center}
\caption{Probability density integrated over the in-plane coordinates of the ground state and a resonant state in $z$ polarization (for $k_z=0$ and MD-B).}
\label{Zdensity}
\end{figure}

If there were no dot, the addition of AlAs  barriers would lead to the formation of GaAs/AlAs SL subbands. When dots are added, these GaAs/AlAs SL states hybridize with the InAs/GaAs SL states, viz. the resonances discussed above. Thus the optical absorption peaks may be viewed as arising from both ``intradot'' (bound-to-resonance) transitions as well as from bound-to-SL continuum transitions. In particular the first continuum state for the $z$ motion of a AlAs-GaAs-WL-dot-GaAs periodic structure is a mixture of states which peak in the middle of the GaAs well (GaAs/AlAs SL) and of states which peak in the WL or nearby the dot. It displays a maximum near the center but also a secondary maximum around the QD/WL with a minimum along z in between (see fig. \ref{Zdensity}). This simplified description helps explaining the smooth evolution of the absorption spectra when going from sample A to sample C. Also it helps explaining the decrease of the integrated absorption from the first peak of sample A to the first peak of sample C as a result of the weakened absorption when one goes from a purely intradot transition (sample A) to transition which involve different spatial localization of the initial state (in the dot) to final state (nearby the dot and in the middle of the SL period).

\section{Experimental results and discussion}

The PL spectra of the MD devices at 5 K shown in fig. \ref{fig:device} exhibit a ground state emission around 1100~meV, while this emission peak gets blue-shifted to 1165meV and 1190~meV for the DD and the SD device respectively. The redshift in sample MD-A is explained by the changed QD growth conditions. The blue-shift of the PL energy of the SD and DD devices in comparison with the MD devices can be explained by the strain distribution in the absence of vertical alignment over more than two growth periods. The vertical alignment of the QDs is driven by the minimization of the free energy of the InAs QDs and the corresponding WL. Consequently, the free energy per QD will be lower in the case of MD devices than for DD and SD devices, and thus the strain induced by the QDs is lower for MD devices. Thus, this strain will lift the energy levels of the QD system \cite{Tadic02}, and the increase of PL energy in the order of MD, DD and SD devices can be qualitatively explained with a corresponding rise of the strain induced by the QDs. Since the size distribution in the QD ensembles causes PL broadening of at least 40~meV, higher excited states of this dot overlap with the ground state emission of small dots peak and will be found at roughly 80~meV higher energies in the PL spectra. 

\begin{figure}[!htbp]
\begin{center}\includegraphics[width=8.5cm,
  keepaspectratio]{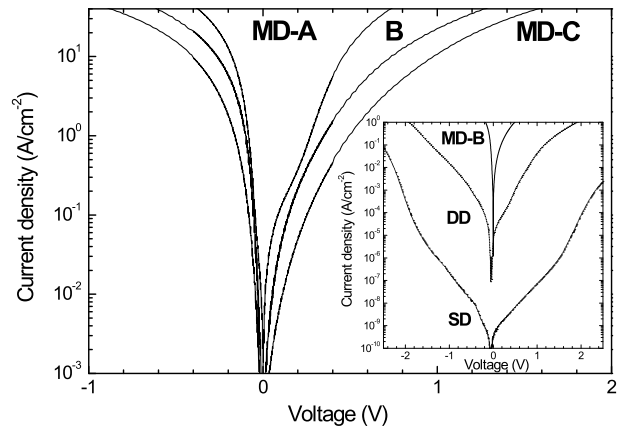}
\end{center}
\caption{Current-voltage (IV) characteristics of devices MD-A, MD-B, and MD-C (Inset: MD-B, DD, SD). All devices
show an asymmetric diode like behavior. The IV shape changes drastically between MD, DD and SD devices.} \label{fig:iv}
\end{figure}

Fig. \ref{fig:iv} exhibits the IV characteristic of the devices at 4 K on a semi-logarithmic scale showing an asymmetric diode-like behavior. We observe that the dark current for the MD devices decreases by one order of magnitude in descending order of MD-A, MD-B and MD-C. An even larger decrease of the dark current occurs in the order of MD-B, DD and SD. As expected, a dark current decrease is observed for the samples with AlAs barriers, which form the SL. In comparison to MD-A the electrons in MD-B moving in the first miniband have to tunnel through the AlAs barriers of the SL. For device MD-C the energetic position of the corresponding miniband is lower compared to MD-B, for which reason the electrons in MD-C have to tunnel through a higher barrier. As a result the dark current decreases in the order of MD-A, MD-B and MD-C. This qualitative description is supported by the decreasing widths of the electronic subbands: $\Delta=100{\rm~meV}$ for MD-A, $\Delta=30{\rm~meV}$ for MD-B and $\Delta=5{\rm~meV}$ for MD-C.

We attribute the drastic decrease for the DD and SD devices to a decreasing influence of the dots for electronic transport within the SL. The strict periodicity of the MD devices is broken for the SD and DD devices. Even if the AlAs barriers continue to be arranged with a period of 11~nm, single or double dot layers are now spaced by 33 and 44~nm, respectively. Our calculations underline the fact that the SL cannot be treated independently from the dots, i.e. if the strict periodicity of the SL with one dot per period is broken we will find a system, where some periods are modified by the dots, while the periods without dots behave rather like a normal one dimensional SL with different energetic properties. Thus we will have a vertical misalignment of the quasi-continuum states, i.e. the minibands, leading to reduced dark currents as well as to lower photocurrents.

\begin{figure}[!htbp]
\begin{center}\includegraphics[width=8.5cm,
  keepaspectratio]{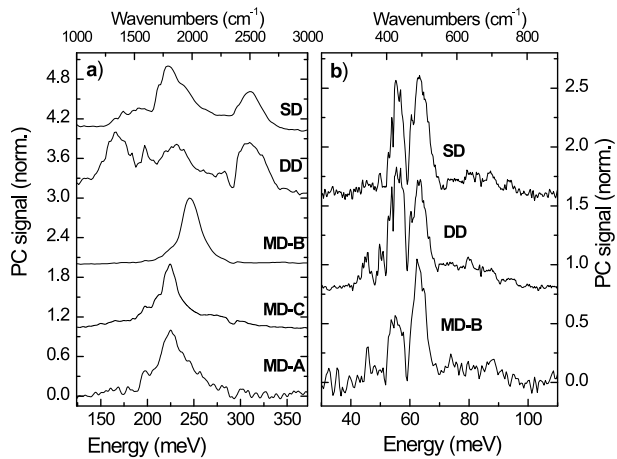}
\end{center}
\caption{a) Photocurrent spectra in the mid-infrared region. b) Far-infrared spectra: The double peak structure
is attributed to the shape asymmetry of the QDs.} \label{fig:pcir}
\end{figure}

Fig. \ref{fig:pcir} shows the PC spectra of MD devices at 4 K, which are characterized by a single main peak at around 225~meV (device MD-A, MD-C) and at 247~meV (device MD-B). The width of these peaks amounts approximately to 40~meV for devices A, B and C, respectively. As theoretically investigated above, this transition can be labeled as ionisation energy of the QD ground states to either the continuum or the blue shifted quasi-continuum of the SL structure. The broadening represents the size distribution of the QD ensemble, to the extend that this broadening is much larger than the auto ionisation width of the resonance. Without vertical coupling of the QDs the situation becomes more complex and several PC peaks in the energy range between 170~meV and 320~meV can be observed (fig. \ref{fig:pcir}a DD and SD). This is not surprising, because the QDs distort quite heavily the energy scheme of the the SL structure periods with dots as compared to the periods without dots. Thus, electron transport along the growth direction becomes more difficult, consequently decreasing the photocurrent signal. Due to the changing energetic structure from periods with dots to SL structure periods without dots and the related miniband split up, additional transitions occur, which are beyond the scope of our theory. However, the change from the single peak PC signal in the MD case to the multipeak signal in the SD and DD devices underlines the strong influence of dots within a SL structure. 
 
Additionally, for the second sample set the energy region around 60~meV was investigated under normal incidence illumination. From the computational approach we would expect an $x$ or $y$ polarization dependent absorption in this energetic region for a bound-to-bound transition, which is equivalent of the $S-P$ atomic absorption. The shape anisotropy, which is responsible for the polarization dependence, causes additionally an energy difference between the $x$ and the $y$ transition. The PC measurement exhibits such a double peak structure with an energy separation of approximately 7~meV in all three devices as shown in fig. \ref{fig:pcir}b. It is important to state that the PC signal is very weak in this measurement since the photo-electrons can only contribute to the signal if they can escape from the excited bound state by thermal activation. However, the double peak structure could be confirmed by transmission experiments on a sample similar to the SD device. The striking feature in this measurement is the homogeneity of the absorption peaks in all three samples. Despite the differences in the interband transitions and PL linewidths above 40~meV, the intersublevel transition can be found at the same spectral position and shows broadening of only 5~meV \cite{Isaia02, Lorke90}.

\begin{figure}[!htbp]
\begin{center}\includegraphics[width=8.5cm,
  keepaspectratio]{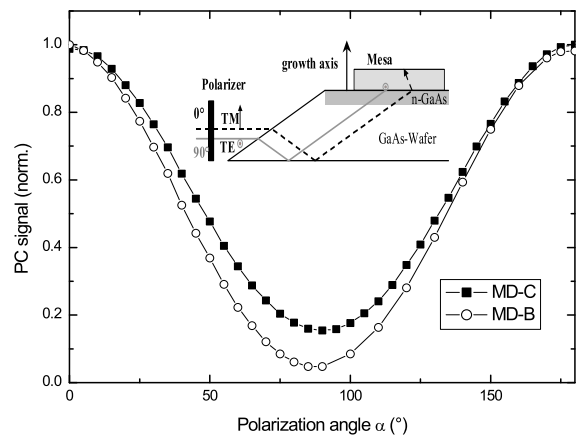}
\end{center}
\caption{Polarization dependence of the photocurrent peak maximum. The inset shows the waveguide structure of
the device with corresponding polarization axes.} \label{fig:polaexp}
\end{figure}

Since the calculations allow predictions about the polarization dependence of the mid-infrared absorption we studied the ionization transition at 250~meV for MD-B under TM and TE polarization, i.e. TM is polarized parallel to the growth direction, while TE corresponds to normal incidence. Fig.\ref{fig:polaexp} exhibits a clear preference of the absorption for the TM illumination like predicted by our calculations. Since a small part of the oscillator strength is left for the TE absorption, the degree of polarization does not drop to zero for this polarization. The TM illumination carries a small share of TE polarization due to the chosen waveguide geometry. These TE shares together with the noise properties of the setup explain why the measured photocurrent at a polarization angle of $90^\circ$ is roughly two times larger than predicted by the theory.

Finally, the suitability of the various device architectures for infrared photodetectors should be briefly discussed. Despite all devices are able to detect infrared light in normal incidence geometry, with or without SL, TM geometry for the ionization transition dominates clearly the PC signal in our devices. Our calculations show that this property is introduced by the QDs themselves and not by the SL. However, these calculations indicate that transitions from a QD ground state to the next excited QD state prefer in-plane polarized light, i.e. normal incidence. Using dots with a bound ground state but excited states in the SL quasi-continuum should lead to devices with strong PC signals under normal incidence. In comparison with SD and DD devices, there is a clear preference for MD devices. Although SD and DD devices have extremely low dark currents, their PC spectrum is complex, difficult to predict and of comparatively low intensity. Also, the PC spectra of MD devices are characterized by an intense single peak, which position can be easily tuned by changing the SL parameters. 

\section{Conclusions}

In summary, we presented an advanced method to design QD based devices by combining the self-organized growth of QD with band gap engineering. In detail, infrared photodetectors made of either vertically stacked InAs QDs or isolated single and double QDs embedded in an AlAs/GaAs SL were realized. The spectral response was predicted and modified by changing the transition energy from the QD ground state to the lowest miniband of the SL. The comparison between QD stacks and isolated QD shows a clear preference of QD stacks concerning the PC intensity and the possibility to design and to tune the spectral dependence of the infrared absorption.

\begin{acknowledgments}
We would like to acknowledge the financial support by the Austrian Science Foundation FWF (SFB ADLIS, START Y47) and the European Community-IST project SUPERSMILE. The LPA-ENS is ``laboratoire associ\'e au CNRS et aux Universit\'es Paris 6 et Paris 7''. One of us (G. B.) acknowledges the Wolfgang Pauli foundation for support.
\end{acknowledgments}

\end{document}